\documentstyle[emulateapj, apjfonts, psfig]{article}

\def\etal{{\it et~al.}}
\def\amin{\ifmmode^{\prime}\else$^{\prime}$\fi}
\def\asec{\ifmmode^{\prime\prime}\else$^{\prime\prime}$\fi}

\def\simgt{\lower.5ex\hbox{$\; \buildrel > \over \sim \;$}}
\def\simlt{\lower.5ex\hbox{$\; \buildrel < \over \sim \;$}}

\newcommand\rcw{\hbox{RCW\hspace{1.5pt}103}}
\newcommand\src{\hbox{1E\hspace{1.5pt}161348$-$5055}}
\newcommand\axs{\hbox{AXS\hspace{1.5pt}J161730$-$505505}}

\newcommand\ASCA{{\it ASCA}}
\newcommand\asca{{\it ASCA}}

\newcommand\ginga{{\it GINGA}}

\newcommand\sax{{\it SAX}}
\newcommand\xte{{\it XTE}}
  
\slugcomment{Draft of Mar 10 2000}

\lefthead{Torii \etal}
\righthead{A Giant Glitch in \axs}

\begin{document}

\title{A Giant Glitch in the Energetic 69 ms X-ray Pulsar \axs}

\author{Ken'ichi Torii$^1$, E.V. Gotthelf$^2$, G. Vasisht$^3$, 
T. Dotani$^4$, \&\ K. Kinugasa$^5$}

\altaffiltext{1}{Space Utilization Research Program (SURP), Tsukuba Space Center (TKSC), National Space Develop Agency of Japan (NASDA), 2-1-1 Sengen, Tsukuba, Ibaraki 305-8505, Japan; torii.kenichi@nasda.go.jp}
\altaffiltext{2}{Columbia Astrophysics Laboratory, 550 West 120th Street, New York, NY 10027; evg@astro.columbia.edu}
\altaffiltext{3}{Jet Propulsion Laboratory, California Institute of Technology, 4800 Oak Grove Drive, Pasadena, CA, 91109; gv@astro.caltech.edu}
\altaffiltext{4}{Institute of Space and Astronautical Science, 3-1-1 Yoshinodai, Sagamihara, Kanagawa 229-8510, Japan; dotani@astro.isas.ac.jp}
\altaffiltext{5}{Gunma Astronomical Observatory, 6860-86 Nakayama, Takayama, Agatsuma, Gunma 377-0702, Japan; kinugasa@astron.pref.gunma.jp}

\begin{abstract}

We present new results on the recently discovered 69 ms X-ray pulsar
\axs, the sixth youngest sample of all known pulsars. We have
undertaken a comprehensive X-ray observing campaign of \axs\ with the
\asca, \sax, and \xte\ observatories and follow its long term
spin-down history between 1989 and 1999, using these, archival \ginga\
and \asca\ data sets, and the radio ephemeris.  The spin-down is not
simply described by a linear function as originally thought, but
instead we find evidence of a giant glitch ($|\Delta P/P|
\stackrel{>}{_{\sim}} 10^{-6}$) between 1993 August and 1997
September, perhaps the largest yet observed from a young pulsar. The
glitch is well described by steps in $P$ and ${\dot P}$ accompanied by
a persistent ${\ddot P}$ similar to those in the Vela pulsar.  The
pulse profile of \axs\ presents a single asymmetric peak which is
maintained over all observation epochs. The energy spectrum is also
steady over time, characterized by a highly absorbed power-law with a
photon index $\Gamma = 1.4 \ \pm\ 0.2$, consistent with that found for
other young rotation powered pulsars.

\end{abstract}


\section{Introduction}

Radio pulsars are thought to be highly magnetized ($ \sim 10^{12}$ G),
rapidly spinning neutron stars whose luminosity is powered by
rotational energy loss.  The study of young ($<10^{5}$ yrs)
rotation-powered pulsars provides an important laboratory for understanding
the early evolution (thermal, spin, and magnetic) of these embers of
stellar collapse.  The most energetic of these pulsars are observable
at X-ray wavelengths, which allow us to probe these extreme, but rare,
examples.

In addition to uniform spin-down corresponding to the rotational
energy loss, rotation powered pulsars show sudden discontinuities in
their rotation periods (see Lyne \& Graham-Smith 1998, chapter 6). These
rare phenomena, known as ``glitches'' are considered to arise from
sudden changes in the configuration of super-dense material in the
neutron star interior. To date, a total of 71 glitches with $|\Delta
P/ P| > 10^{-9}$ have been reported in 30 pulsars (Urama \& Okeke
1999). Observation of pulsar glitches gives us insights into the
structure and physical processes inside the neutron stars, such as
the interactions of neutron superfluid and crust components (e.g., Anderson
\& Itoh 1975).

In this article we report the detection of a giant glitch from \axs\
by using the newly acquired multi-mission X-ray data. This source
 is an unusual case of a young pulsar discovered first by its
X-ray emission, revealed during the course of an archival X-ray study
of the SNR \rcw\ (Gotthel, Petre, \& Hwang 1997). Further analysis
detected highly significant pulsations from photons attributed to this
source (Torii \etal\ 1998). The 69 ms pulse period was consistent with
that reported from a \ginga\ observation of the region (Aoki, Dotani,
\& Mitsuda 1992). Recent radio observation has confirmed \axs\ as a
young energetic rotation-powered pulsar (Kaspi \etal\ 1998). The lack
of evidence for this object in archival soft X-ray images of this
field suggested that the source might be highly absorbed, making it
difficult to determine whether it is associated with a SNR.

\section{Observations}

A set of day-long X-ray observations of the field containing \rcw,
\src, and \axs\ were performed with the \asca\ (Tanaka, Inoue, \& Holt 1994),
\sax\ (Boella \etal\ 1997), and \xte\ (Bradt \etal\ 1993)
observatories. We summarize in table 1 the set of observations
presented in this work. 

Imaging data were acquired with the Gas Imaging
Spectrometers (GIS) on-board \ASCA\ and with the Medium Energy
Concentrators (MECS) instruments on \sax.  These instruments have
moderate imaging ($\sim 2^{\prime}$) and spectral resolution ($\sim
8\%$ at 6 keV) over an energy band pass of $0.7-10$~keV (GIS) and
$1.5-12$~keV (MECS),
with a field-of-view (FOV) large enough to cover the SNR and pulsar 
simultaneously. Non-imaging data were obtained with the
Proportional Counter Array (PCA) on-board \xte\ which provides broader
energy band-pass ($2-60$ keV) at lower spectral resolution ($\sim
16\%$ at 6 keV). The PCA FOV is roughly circular with a
$\sim 1^{\circ}$ FWHM response. 
 The GIS data were collected in the highest time resolution
mode ($0.5 \ \rm{ms}$ or $61 \ \mu\rm{s}$, depending on data
acquisition mode) whose measured absolute accuracy is $200 \
\mu\rm{s}$ in this mode (Saito \etal\ 1997). The PCA data were
collected using the Good Xenon mode  with $0.9
\mu$s timing resolution.  For the current analysis, the absolute
timing uncertainty is $\sim 100 \mu$s (Rots \etal\ 1998). Photons
collected by the MECS are time tagged with 
15~$\mu$s resolution. 
We do not include data from \asca 's
Solid-state Imaging Spectrometers (SIS), as the pulsar fell just off
the edge of its FOV. Nor do we include data from \sax 's other
instruments as the observing time is insufficient for these
instruments to measure the pulsar periodicity reliably, as a part of the observation was
interrupted prematurely.

Each data set was processed through its standard pipeline reduction
for that mission and edited to exclude times of high background
contamination using the standard screening criteria. This rejects time
intervals of South Atlantic Anomaly passages, Earth occultations,
bright Earth limb in the FOV (\asca\ and \sax\ only), and other
periods of high particle activity. The resulting effective observation
times are summarized in Table 1. For each observation, event data from
all detectors were co-added and the arrival times of each event were
corrected to the solar system barycenter using the software {\tt
TIMECONV} (\asca), {\tt BARYCONV} (\sax), or {\tt FXBARY} (\xte).

\section{Results}

\subsection{Timing}

The X-ray images obtained with both \asca\ and \sax\ above 3~keV
reveal AXS~J161730--505505 $4'$ outside the SNR shell (see figure 1a
and 1b of Gotthelf \etal\ 1997; Gotthelf, Petre, \& Vasisht 1999). To increase the signal-to-noise
ratio for detecting pulsations from the pulsar, we extracted photons
from an 8$'$ diameter aperture centered on the pulsar, restricting the
energy range of extracted photons to $3-10$ keV for GIS and $3-12$ keV
for the other instruments.  For the PCA data, in this energy band, we
further restrict our search to Layer 1 data only, which provides the
best sensitivity for a Crab-like spectrum; For the higher energy
analysis afforded by the PCA, above $\sim 12$ keV, we used data from
all three PCA layers.

We searched each data set for the expected 69 ms period predicted from
the initial period and period derivative measurement (Torii \etal\
1998).  A periodgram was constructed using the $\chi^2$ statistic to
test against a null hypothesis.  For each trial period, we folded the
data into 10 bins and computed the $\chi^2$ of the resultant
profile. We search a narrow range of periods centered on the expected
period $\pm 0.1$~ms, sampled in increments of $0.1 \times P^2/T$,
where $T$ is the observation duration, and $P$ is the test period.  A
highly significant signal was detected from each of our data sets.

As well as the newly obtained data, we have re-analyzed the previous
\ginga\ and \asca\ data (Aoki \etal\ 1992; Torii \etal\ 1998) in a
uniform way and revised the period and its error by using the method
of Leahy (1987). Our X-ray timing results derived from these 13
measurements are listed in table 1, along with an updated radio
ephemeris (Kaspi 1999, Private communication).

\subsection{Spectrum}

We search for spectral dependence of the pulse profile by comparing the
folded light curves in several energy bands. No strong energy
dependence is evident in the energy resolved light curves.  Furthermore, 
the pulse amplitude and pulse profile remained unchanged between
observational epochs.  

We examined the \asca\ and \sax\ data on \axs\ for any long term
changes in its energy spectrum or flux. As for the timing analysis, we
restrict our comparison to the energy range above $3$ keV and extract
photons from an $8'$ diameter aperture centered on the source. We
fitted the spectrum with a power law function modified by interstellar
absorption. The absorption was fixed at $6.8\times 10^{22} \ {\rm
cm^{-2}}$ (Torii \etal\ 1998). Spectra from each observation were
found to be consistent with each other.
Combining the 7 \asca\ observations, we obtain the pulse phase
averaged photon index, $\Gamma = 1.4\pm0.2$, and the observed flux of
$(3.6\pm 0.2)\times 10^{-12}\, {\rm ergs\,s^{-1}\,cm^{-2}}$ (90\%
confidence errors) in the
3-10~keV range, which is consistent with the previous measurement
(Torii \etal\ 1998).

\section{Discussion}

A $\chi ^2$ fit to the all 14 data points as summarized in table 1
gives the mean spin-down rate of 
$\dot{P}= 1.3611(1) \times 10^{-13}$~s/s and $P=0.069347150(1)$
at MJD 50,000.0 (Figure 1). However,
the quality of the fit is bad, with 
$\chi ^2/d.o.f. = 4808/12$. The residual of the fit shows a jump of
$\Delta P \simeq -1.2\times 10^{-7}$~s between the observations of
1993 August (MJD 49,217.6) and 1997 September (MJD 50,696.0) (Figure
2, top panel).  Within the observation span of 10 years, the residual
is neither periodic nor smooth. A sudden change in the period between
MJD 49,217.6 and MJD 50,696.0 is suggested. We consider the most
likely explanation for these residuals is due to a glitch, similar to
those observed in several young rotation-powered pulsars.

Given the clear evidence of glitch activity, we next attempted to model
the spin-down data with a single glitch followed by an exponential
recovery. The data coverage is limited and we simply assumed the
following relation for the spin-down.
\begin{equation}
P(t) = P_0 + {\dot P} \cdot (t-t_0) + \Delta P \cdot \exp (-\frac{t-t_0}{\tau})
\end{equation}
where $\Delta P=0$ for $t<t_0$ and $\Delta P$ is a negative constant
for $t\geq t_0$. This model contains five parameters, which, except
for the depth of the glitch $\Delta P$, are found to be independent of
the time of the glitch, $t_0$. The derived parameters, $P_0$, ${\dot
P}$, $\Delta P$, and $\tau$ are summarized in table 2 for assumed
values of $t_0 = 49,300.0$~MJD, $t_0 = 50,000.0$~MJD, and $t_0 =
50,600.0$~MJD. The residual for $t_0=50,000$ is shown in the middle panel of figure
2. The quality of the fit is now characterized by $\chi^2 /d.o.f. =
22.6/10$.

The size of the glitch
depends strongly upon the unknown glitch epoch $t_0$. The fractional increase in rotation was found to be $\Delta
P/P=-4.2\times 10^{-6}$ for $t_{0}=50,000.0$~MJD (fixed) while it
changes between $\Delta P/ P=-11\times 10^{-6}$ for $t_0=49,300.0$~MJD
(fixed) and $\Delta P/P=-1.8\times 10^{-6}$ for $t_0=50,600.0$~MJD
(fixed). The minimum
fractional increase in rotation rate is therefore comparable with those of
the  largest known pulsar glitches (Lyne, \etal\
1996a; Shemar \& Lyne 1996).

Using the above model, the recovery time following the glitch episode
is found to be $\tau=700$~days. This duration is somewhat unusual
compared to the radio pulsars whose recovery time is seen to bifurcate
between $\tau \simeq 100$~days and $\tau \geq 1,000$~days (Shemar \&
Lyne 1996). 
Our derived value lies squarely between these two time scales, perhaps 
due to the simple model we invoked which allows for only a single glitch.
Because of the sparse data coverage between
August 1993 and September 1997, however, we cannot determine if the
recovery time could be expressed as the sum of the two timescales.

Glitches found in radio pulsars may be classified into three groups
(Lyne \& Graham-Smith 1998). The first is a Crab-like glitch which is
characterized by steps mainly in ${\dot P}$. The second is a Vela-like
glitch which is characterized by large changes in $P$ ($\Delta P/P
\simeq -10^{-6}$) and the exponential recoveries. For the Vela pulsar,
linear sawtooth changes in ${\dot P}$ have been observed between glitches (Lyne
\etal\ 1996b). The third kind is often found in old pulsars which is
characterized by a change in $P$.

Since the large glitch found for \axs\ is similar in its magnitude to
those in the Vela pulsar, a phenomenological model taking into account
the sawtooth behavior may be a good description. Apart from short term 
effects, the spin-down for the Vela pulsar is expressed by a linear
change in spin-down rate. Therefore, the following function is
appropriate if the transient effects have already ceased by MJD 50,696.
\begin{equation}
P(t) = P_0 + \Delta P_0 + ({\dot P}+\Delta {\dot P}) \cdot
(t-t_0) + \frac{1}{2}{\ddot P} \cdot (t-t_0)^2
\end{equation}
Here, $\Delta P_0=0$ and $\Delta {\dot P}=0$ for $t < t_0$ and they
are constant values ($\Delta P_0 <0$ and $\Delta {\dot P}>0$) for $t
\geq t_0$. This model contains six parameters. Again, the time of the
glitch, $t_0$, had to be assumed. The derived parameters are
summarized in table 3. For the condition that $\Delta P_0 < 0$, $t_0$
was restricted to $t_0 \stackrel{<}{_\sim} 50,205$.  The residual for $t_0=50,000$ is
shown in the bottom panel of figure 2.  The quality of the fit is
significantly improved to $\chi^2 /d.o.f. = 17.1/9$. Compared to the
fit to Equation (1), the f-test gives
a chance probability of $\sim 0.1$. This result suggests that
transient effects had indeed ceased by MJD~50,696. The reduced $\chi
^2$ is still larger than unity, suggesting the presence of timing 
noise and smaller glitches.

We can estimate the expectancy of large ($|\Delta P/P| > 10^{-7}$)
glitches by using the semi-empirical relation based upon the
superfluid vortex unpinning model (Alpar \& Baykal 1994).  Using this
relation, the expected number of large glitches between the first
\ginga\ observation and the last \asca\ observations is
3.7. Therefore, there should have been about 4 glitches of $|\Delta
P/P| > 10^{-7}$. Indeed the residuals to the fit of Equation (1) or
(2) still hint a small jump of $\Delta P = -5\times 10^{-9}$~s between
MJD~51,263.8 (1999 Mar.) and MJD~51,394.2 (1999 Aug). This may be
another glitch of $\Delta P/P= -7\times 10^{-8}$, much smaller in
magnitude than the one near MJD~50,000, but still relatively large
compared to those seen for most radio pulsars (Shemar \& Lyne 1996).  We
have thus found a giant glitch of $|\Delta P/ P| \stackrel{>}{_{\sim}}
10^{-6}$ at $t_0 \simeq 50,000$~MJD ($49,218 \leq t_0 \leq 50,696$)
and possibly a glitch of $|\Delta P/ P| \simeq 7\times 10^{-8}$ at $t_0
\simeq 51,300$~MJD ($51,264 \leq t_0 \leq 51,394$).

 The detection of a giant glitch from \axs\ gives a rare sample for
studying the origin of pulsar glitches and the interior structure of
neutron stars. In this context, regular timing observation of this
pulsar in the radio band is quite important to monitor the onset of a
glitch and following transient effects. Long term timing observations
for measuring the braking index are desired. Searches for the highly
absorbed X-ray emission from the supernova remnant associated with
this pulsar should be undertaken with the next generation X-ray
observatories.

\begin{acknowledgements}
{\noindent Acknowledgments} --- 

We thank Jules Halpern for a critical reading of the manuscript, the
referee Andrew Lyne for invaluable comments and suggestions on the
original draft, and Victoria Kaspi for communicating the revised radio
ephemeris. Special thanks to Alan Smale for expediting the delivery of the
\xte\ data and to the XTEhelp@athena.gsfc.nasa.gov crew for their patience
and assistance.  Part of this research has made use of data obtained
through the HEASARC data center, provided by NASA/GSFC. E. V. G. and G. V.
acknowledge the supported of NASA LTSA grant NAG~5-7935.

\end{acknowledgements}

\clearpage

\begin{deluxetable}{lllllll}
\small
\tablewidth{0pt}
\tablecaption{Timing Data for \axs. \vfill
\label{table I}}
\tablehead{
\colhead{\hfil Number \hfil}&\colhead{\hfil Satellite/ \hfil} &\colhead{\hfil Instrument \hfil} &
\colhead{Obs Date${\rm ^a}$}& \colhead{Exposure} & \colhead{Epoch${\rm
^b}$} & \colhead{Period${\rm ^c}$} \nl
 \colhead{}  & \colhead{Observatory}& \colhead{} & \colhead{} & \colhead{(s)} & \colhead{(MJD)} & \colhead{(ms)} \nl
}
\startdata
\noalign{\vskip 5pt}
1&\ginga	& LAC 	& 1989 Mar 4  	& 14,400 & 47590.2 & 69.31890(3) \nl
2&\asca 	& GIS  	& 1993 Aug 17 	& 35,353 & 49217.277625 & 69.338019(2)\nl
3&\asca 	& GIS  	& 1997 Sep 4  	& 58,063 & 50696.736757 &
69.355301(1) \nl
4&\xte  	& PCA  	& 1998 Jan 2  	& 33,712 & 50816.132391 & 69.356717(1) \nl
5&Parkes  	& 64~m & 1998 Jan 15 & --- & 50829.7 &
69.356889(7)${\rm ^d}$ \nl
6&\sax  	& MECS 	& 1998 Sep 17 	& 16,288 & 51073.892337 & 69.359778(1) \nl
7&\asca	& GIS	& 1999 Feb 12	& 15,405 & 51222.279184 & 69.361532(4) \nl
8&\asca	& GIS	& 1999 Feb 20 	& 16,232 & 51229.708310 & 69.361624(7) \nl
9&\xte  	& PCA  	& 1999 Mar 5  	&  37,253 & 51243.247065 & 69.3617824(5) \nl
10&\asca	& GIS	& 1999 Mar 6	& 20,700 & 51243.354377 & 69.361782(8)\nl
11&\sax & MECS 	& 1999 Mar 23	& 38,197 & 51261.461543 & 69.362000(3)\nl
12&\asca	& GIS	& 1999 Mar 26	& 18,914 & 51263.519288 & 69.362021(3) \nl
13&\sax	& MECS	& 1999 Aug 4	& 53,115 & 51394.834928 & 69.363568(2) \nl
14&\asca	& GIS	& 1999 Sep 13	& 17,796 & 51434.716949 & 69.364037(6) \nl
\enddata
\tablenotetext{a} {Start time of the observation.}
\tablenotetext{b} {Middle time of the observation except for the radio 
ephemeris.}
\tablenotetext{c} {1$\sigma$ error of the last significant figure is
shown in parentheses.}
\tablenotetext{d} {The radio ephemeris reported here (Kaspi 1999,
Private communication) has been revised from that given in Kaspi \etal\ (1998).}
\end{deluxetable}

\begin{deluxetable}{lllllll}
\small
\tablewidth{350pt}
\tablecaption{Best-fit Parameters of a Giant Glitch in \axs\ for Equation 
(1). \vfill
\label{table II}}
\tablehead{
\colhead{\hfil Parameter \hfil}& &\colhead{\hfil Values \hfil} & \nl
}
\startdata
$t_0$ [MJD] & 49,300 (fixed) & 50,000 (fixed) & 50,600 (fixed)\\
$P_0$ [s]$^a$  & 0.069338991(2) & 0.069347220(3) & 0.069354272(5)\\
${\dot P}$ [s/s]& $1.3605(5)\times 10^{-13}$ & $1.3605(5) \times 10^{-13}$ &$1.3605(5) \times 10^{-13}$ \\
$\Delta P$ [s]& $-8(2)\times 10^{-7}$& $-2.9(2) \times 10^{-7}$ & $-1.26(4)\times10^{-7}$\\ 
$\tau$ [days] & $7(1)\times 10^2$ & $7(1)\times 10^2$ & $7(1)\times 10^2$\\ 
\enddata
\tablenotetext{a} {The period just before the glitch at $t_0$.}
\end{deluxetable}

\begin{deluxetable}{lllllll}
\small
\tablewidth{350pt}
\tablecaption{Best-fit Parameters of a Giant Glitch in \axs\ for Equation (2). \vfill
\label{table III}}
\tablehead{
\colhead{\hfil Parameter \hfil}& &\colhead{\hfil Values \hfil} & \nl
}
\startdata
$t_0$ [MJD] & 49,300 (fixed) & 50,000 (fixed) & 50,205 (fixed)\\
$P_0$ [s] & 0.069338980(3) & 0.06934707(3) & 0.06934943(3)\\
$\Delta P_0$ [s] & $-4.6(3)\times 10^{-7}$ & $-1.0(2)\times10^{-8}$ & $-3.1(3)\times10^{-10}$\\
${\dot P}$ [s/s]& $1.344(3)\times 10^{-13}$ & $1.332(5) \times10^{-13}$ &$1.328(5) \times 10^{-13}$ \\
$\Delta {\dot P}$ [s/s]& $5.8(7)\times 10^{-15}$ & $5.8(7)\times 10^{-15}$ &$5.8(7) \times 10^{-15}$ \\
${\ddot P}$ [s$^{-1}$]& $-2.0(3)\times 10^{-23}$& $-2.0(3)\times 10^{-23}$ & $-2.0(3)\times10^{-23}$\\ 
\enddata
\end{deluxetable}

\begin{figure}
\centerline{
\psfig{figure=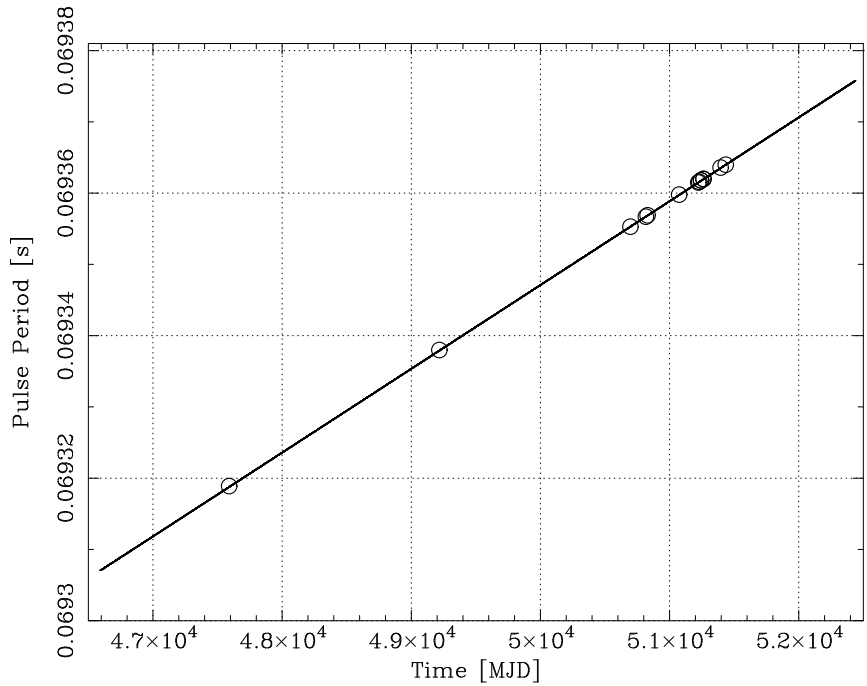,height=8.0cm,angle=270} 
}
\bigskip
\caption{The pulse period evolution of \axs\ over 10 years. }
\end{figure}

\begin{figure}
\centerline{
\psfig{figure=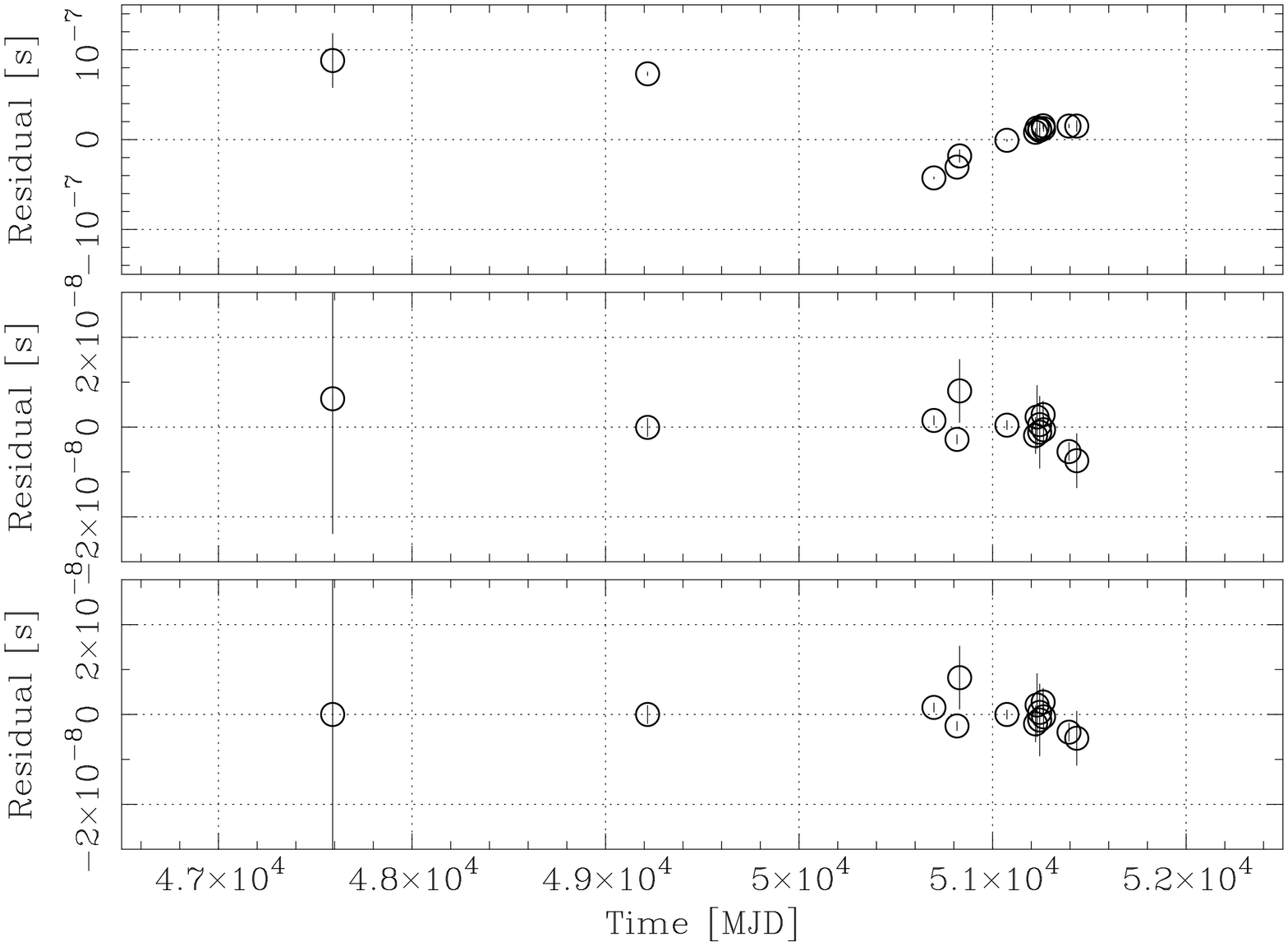,height=8.0cm,angle=270} 
}
\bigskip
\caption{{\it Top Panel)} Pulse period residual as fitted by a linear
function. {\it Middle Panel)} Pulse period residual as fitted to
Equation (1) with a glitch at MJD 50,000. {\it Bottom Panel)} Pulse
period residual as fitted to Equation (2) with a glitch at MJD 50,000.}
\end{figure}


\begin{references}
\reference{Alpar1994} Alpar, M.A., \& Baykal, A. 1994, MNRAS, 269, 849
\reference{Anderson1975} Anderson, P.W., \& Itoh, N. 1975, Nature,
256, 25
\reference{Aoki1992} Aoki, T., Dotani, T., \& Mitsuda, K. 1992, IAU Circ., 5588
\reference{Boella1997} Boella, G., Butler, R.C., Perola, G.C., Piro,
L., Scarsi, L., \& Bleeker, J.A.M. 1997, A\&AS, 122, 299
\reference{Bradt1993} Bradt, H.V., Rothschild, R.E., \& Swank,
J.H. 1993, A\&AS, 97, 355
\reference{Gotthelf1997} Gotthelf, E.V., Petre, R., \& Hwang, U. 1997,
ApJ, 487, L175
\reference{Gotthelf1999} Gotthelf, E.V., Petre, R., \& Vasisht, G. 1999, ApJ, 514, L107
\reference{} Kaspi, V.M., Crawford, F., Manchester, R.N., Lyne, A.G., 
Camilo, F., D'amico, N., \& Gaensler, B.M. 1998, ApJ, 503, L161
\reference{} Leahy, D.A. 1987, A\&A, 180, 275
\reference{Lyne1996a} Lyne, A.G., Kaspi, V.M., Bailes, M., Manchester,
R.N., Taylor, H., \& Arzoumanian, A. 1996a, MNRAS, 281, L14
\reference{Lyne1996b} Lyne, A.G., Pritchard, Graham-Smith, F., \&
Camilo, F. 1996b, Nature, 381, 497
\reference{Lyne1998} Lyne, A.G., \& Graham-Smith, F. 1998, Pulsar
Astronomy Second Edition (Cambridge: Cambridge Univ. Press)
\reference{} Rots, A.H., et al. 1998, ApJ, 501, 749
\reference{} Saito, Y., et al. 1997, \asca\ News, 5
\reference{}Shemar, S.L., \& Lyne, A.G. 1996, MNRAS, 282, 677
\reference{Tanaka1994} Tanaka, Y., Inoue, H., \& Holt, S.S. 1994, PASJ, 46, L37
\reference{Torii1998} Torii, K., Kinugasa, K., Toneri, T., Asanuma,
T., Tsunemi, H., Dotani, T., \& Mitsuda, K. 1998, ApJ, 494, L207
\reference{Urama1999} Urama, J.O., \& Okeke, P.N. 1999, MNRAS, 310, 313

\end{references}
\end{document}